\documentclass[%
 preprint, superscriptaddress,
onecolumn,
showpacs,
 amsmath,amssymb,
]{revtex4-1}
\usepackage{mathrsfs}
\usepackage{graphicx}
\usepackage{dcolumn}
\usepackage{bm}
\usepackage{amsmath,mathrsfs}    
\usepackage{graphicx}   
\usepackage{subfigure}  
\usepackage{hyperref}   
\usepackage{upgreek}
\usepackage{textcomp}

\newcommand{\T}[1]{\textnormal{#1}}
\newcommand{\scL}{\ensuremath{\mathrm{\mathscr{L}}}}

\begin{document}
\title{Photon noise suppression by a built-in feedback loop}
\author{A.~Al-Ashouri}
\author{A.~Kurzmann}
 \email{annika.kurzmann@uni-due.de}
\author{B.~Merkel}
\affiliation{University of Duisburg-Essen, Faculty of Physics and CENIDE, Lotharstr. 1, 47057 Duisburg, Germany}
\author{A.~Ludwig}
\author{A.~D.~Wieck}
\affiliation{Lehrstuhl f\"ur Angewandte Festk\"orperphysik, Ruhr-Universit\"at Bochum, Universit\"atsstra{\ss}e 150, 44780 Bochum, Germany}
\author{A.~Lorke}
\author{M.~Geller}
\affiliation{University of Duisburg-Essen, Faculty of Physics and CENIDE, Lotharstr. 1, 47057 Duisburg, Germany}

\date{\today}

\begin{abstract}
Visionary quantum photonic networks need transform-limited single photons on demand. Resonance fluorescence on a quantum dot provides the access to a solid-state single photon source, where the environment is unfortunately the source of spin and charge noise that leads to fluctuations of the emission frequency and destroys the needed indistinguishability. We demonstrate a built-in stabilization approach for the photon stream, which relies solely on charge carrier dynamics of a two-dimensional hole gas inside a micropillar structure. The hole gas is fed by hole tunneling from field-ionized excitons and influences the energetic position of the excitonic transition by changing the local electric field at the position of the quantum dot. The standard deviation of the photon noise is suppressed by nearly 50 percent (noise power reduction of 6 dB) and it works in the developed micropillar structure for frequencies up to 1 kHz. This built-in feedback loop represents an easy way for photon noise suppression in large arrays of single photon emitters and promises to reach higher bandwidth by device optimization.

\end{abstract}

\maketitle

\section{Introduction}
Single, transform-limited photons are an essential ingredient for transmission of quantum information over long distances in a quantum network \cite{Gisin.2002} , where the quantum bits (qubits) \cite{Bernien.2013} are stored in nodes and coupled via single photon channels \cite{Kimble.2008,Flagg.2010}. Self-assembled quantum dots (QDs) are solid-state quantum emitters that produce on one hand antibunched \cite{Muller.2007b,Flagg.2009}, transform-limited indistinguishable photons \cite{Santori.2002,Laurent.2005} and photon pairs \cite{Dousse.2010,Muller.2014} with a high photon flux \cite{Kuhlmann.2015,Matthiesen.2012}. The anti-bunching and indistinguishability can be further enhanced under resonant excitation \cite{Ates.2009,Matthiesen.2012,Matthiesen.2013}. On the other hand, they offer the ability to store the quantum information locally using electron spins \cite{Atature.2006}, and thus providing a building block for stationary qubits as well. Moreover, an important advantage is the convenient integration of quantum dots into functional semiconductor heterostructures, in order to tune the energetic levels of different dots into resonance by electrical \cite{Li.2000,Warburton.2002} or electro-elastic means \cite{Trotta.2012,Kuklewicz.2012}. Arrays of self-assembled QDs with identical emission energies and a spin-photon interface, mapping the spin quantum state on single indistinguishable photons is therefore the basic ingredient for a photonic quantum network \cite{Lodahl.2015}; at best embedded in a nanophotonic waveguide \cite{Sollner.2015,Coles.2016} for an effective coupling between the different nodes.   

However, a major challenge is charge and spin noise in the environment of the quantum dot \cite{Kuhlmann.2013} that introduces decoherence \cite{Fischer.2009} by random frequency and intensity fluctuations of the emitted photons. This destroys the transform-limited linewidth as well as the photon indistinguishability. The spin noise comes from fluctuations of the nuclear spins that interact with the electron spin via hyperfine interaction \cite{Khaetskii.2002} and add up to an effective magnetic field (Overhauser field). It dominates at high frequencies up to 100 kHz \cite{Kuhlmann.2013}. The electrical charge noise shifts the emission wavelength by the quantum-confined Stark effect \cite{Kuhlmann.2013,Kuhlmann.2015}. The charge noise has a 1/f-like power dependence. While its origin is still not clear in detail, charge traps in the environment can lead to jumps and drifts of the emission frequency by the Stark effect \cite{Houel.2012}. Shortening the radiative lifetime by placing the quantum dot into a microcavity (Purcell effect) \cite{Laurent.2005,Gazzano.2013,Somaschi.2016} or weak excitation of the quantum dot in the Heitler regime \cite{Matthiesen.2012,Matthiesen.2013} are just two examples to circumvent the photon decoherence. Another approach is to use a feedback-loop that counteracts the spectral fluctuations. One option is an external feedback that measures the photon frequency and corrects any spectral diffusion by an external applied gate bias, using the Stark effect \cite{Acosta.2012,Prechtel.2013,Hansom.2014} Alternatively, an internal feedback mechanism uses the coupling of the electron-hole pairs to the environment. The exciton coupling to the nuclear spins causes a hysteresis and dragging of the resonance \cite{Hogele.2012, Chekhovich.2013} and can be used to stabilize the emission line \cite{Latta.2009,Yang.2013}.   

In this paper, we present a conceptually simple stabilization mechanism, realized by an internal coupling between a QD and a two-dimensional hole gas (2DHG). The QD fluorescence is stabilized automatically by the built-in dynamics of optical exciton generation and charge transport into the hole gas. This eliminates the need for an external magnetic field or a complicated active feedback system and thus enables stabilization of numerous QDs on the same integrated chip. The observation of this feedback loop is realized by the dot being embedded in a micropillar-patterned, mesa-like heterostructure, which laterally confines the holes generated from a field-ionization process of the resonantly generated excitons. This work demonstrates a photon noise suppression for frequencies up to 1 kHz, where the standard deviation of the noise is reduced by 50 percent to 90 MHz (0.39 $\mu$eV). The photon noise reduction and the maximum stabilization frequency depend on the hole tunneling rates and the lateral hole gas confinement. Optimization of the heterostructure layer sequence and size of the micropillar structure promise even faster stabilization up to the desired 100 kHz, which is desired to cancel out spin and charge noise.

\begin{figure*}
	\centering
	\includegraphics[width=1\columnwidth]{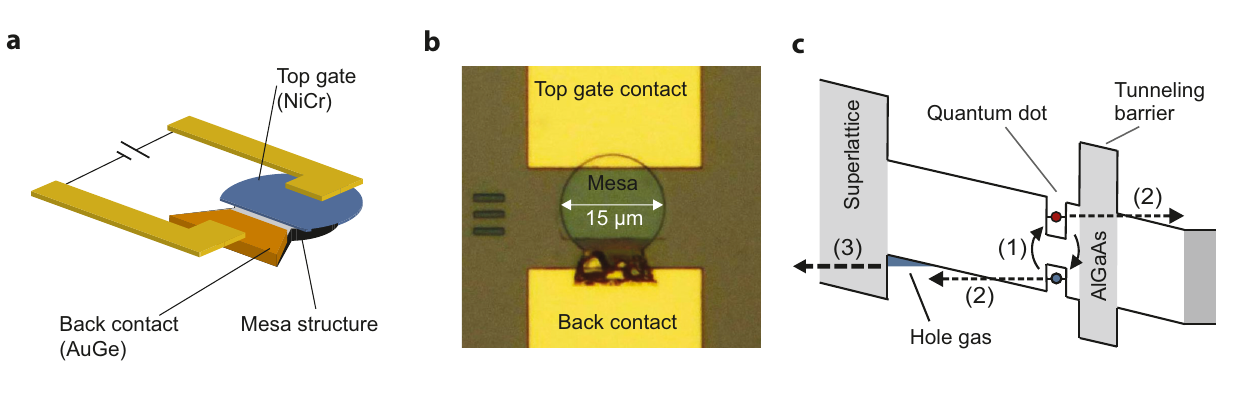}
	\caption{The heterostructure device for confining a two-dimensional hole gas at an AlGaAs/GaAs interface. \textbf{a}, The semiconductor material is processed into a mesa structure with a transparent, 6\,nm thick nickel-chromium top gate (blue). The Ohmic back contact consists of a Si-doped GaAs-layer with an electrical connection, established by annealing an AuGe-alloy (orange). The active area, defined by etching, is a micropillar mesa structure with height of 200\,nm and a NiCr gate on top. \textbf{b}, Optical microscopy image of the device with the 15 $\mu$m mesa structure. \textbf{c}, Schematic diagram of the conduction and valence band edges of the active region. An applied gate voltage tunes the electric field such that optically-generated excitons in the QD can be field-ionized. Arrows indicate the processes involved in the hole gas formation: (1) Exciton generation in the QD, (2) tunneling of electron and hole, (3) and leakage of the hole gas.}
	\label{f:dev}
\end{figure*}

A schematic illustration of the device structure, together with an optical microscope image of the actual device is shown in Fig.~\ref{f:dev}(a) and (b), respectively. The electric field across the device can be controlled by an applied voltage between the Ohmic back contact (orange area in Fig.~\ref{f:dev}(a)) and top gate (blue area in Fig.~\ref{f:dev}(a)). The schematic band structure in Fig.~\ref{f:dev}(c) shows the active part of the heterostructure. It consists of an AlGaAs tunneling barrier between the QD and the n-doped back contact and a GaAs capping layer on top of the dot layer, followed by a GaAs/AlAs superlattice (for more details: methods section). The interface between this superlattice and the capping layer is where the two-dimensional hole gas is formed in growth direction, as depicted in Fig.~\ref{f:dev}(c). In general, charge carriers trapped at interfaces or in deep levels influence the resonance energy of the excitonic transitions by an internal electric field \cite{Seidl.2005,Houel.2012} and the effect of charge accumulation in a two-dimensional hole gas has been observed before \cite{Luttjohann.2005,Bakker.2015}. The stabilization of the photon stream relies here on a feedback loop between the hole gas population and the resonance frequency of the QD, as described in Merkel et al.~\cite{Merkel.2017} before.

\section{Feedback loop}

\begin{figure*}
	\centering
	\includegraphics[width=1\columnwidth]{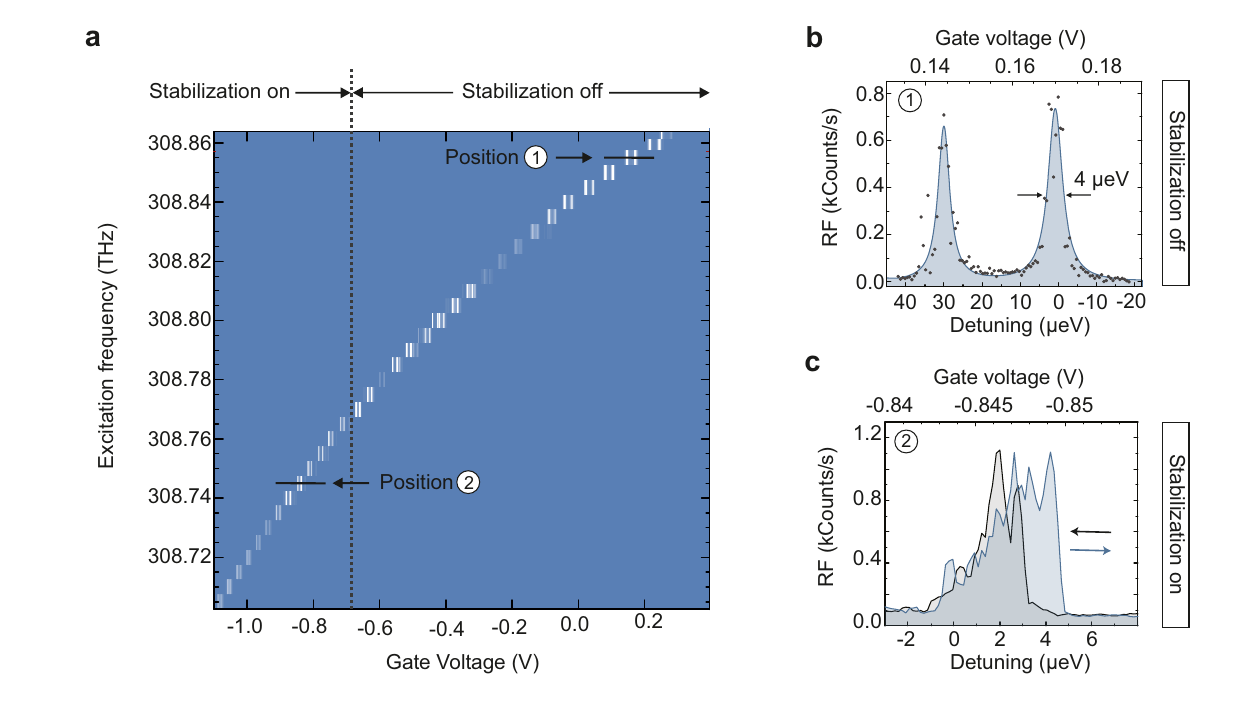}
	\caption{The influence of the hole gas on the resonance fluorescence (RF). \textbf{a}, RF scan of the exciton (X) with its fine structure splitting for different gate voltages and laser excitation frequencies. The dotted vertical line separates the voltage ranges where the stabilization is active or inactive, respectively. \textbf{b}, A line cut at position (1) through the two-dimensional RF scan at weak electric fields, showing the usual excitonic fine-splitting and Lorentzian lineshape at a laser excitation power of $\Omega_\mathrm{R}=0.14\,\Gamma_0$, with $\Gamma_0= 0.97$\,GHz.  \textbf{c}, Line cut at position (2) at a gate voltage of about -0.8\,V on one exciton transition in the hole gas regime, where the stabilization is switched on. A dragging and hysteresis effect emerges, resulting from the interaction between the hole gas and the quantum dot. The laser excitation power in Rabi frequency is here $\Omega_\mathrm{R}=0.2\,\Gamma_0$. }
	\label{f:RF}
\end{figure*}

The voltage applied to the gate contact (cf.~Fig.~\ref{f:dev}) controls externally the electric field at the QD, and the excitonic transitions are shifted via the quantum-confined Stark effect, as shown in  Fig~\ref{f:RF}, where the resonance fluorescence of the exciton (X) is given for different excitation frequencies and applied gate voltages. The Stark-shift of the exciton is visible over an extended voltage range, starting below $V_g=$ -1 V to above 0.2 V, where tunneling into the dot quenches the exciton transition \cite{Kurzmann.2016b}. The resonance fluorescence scan can be divided into two areas, where the stabilization by the hole gas is active (below $V_g=-0.7$ V) or inactive (above $V_g=-0.7$ V). The stabilization is inactive for high negative gate voltages, as the hole escape rate $\gamma_\T{leak}$ (process (3) in Fig.~\ref{f:dev}(c)) always depletes the hole gas completely (see rate equation model in supplemental information in Merkel et al.~\cite{Merkel.2017}). When the stabilization is "off", a typical resonance fluorescence spectrum with a symmetrical Lorentzian lineshape is observed in Fig.~\ref{f:RF}(b). This spectrum was measured for a laser frequency of 308.855 THz and is visible at position (1) in the two-dimensional resonance fluorescence (RF) scan in Fig.~\ref{f:RF}(a). The two fine-structure split excitonic lines with a linewidth of $4 \,\mu$eV (corresponding to $\Gamma_0= 970$\,MHz) are visible at a low laser excitation power (expressed in terms of the Rabi frequency) of about $\Omega_\mathrm{R}=0.14\,\Gamma_0$.

An easy indicator for "stabilization on" is a hysteresis behaviour and a dragging of the resonance with an asymmetric lineshape for gate voltages below $V_g=-0.7$ V (visible in Fig.~\ref{f:RF}(c)). The hole gas builds up at the AlGaAs/GaAs interface and acts as an internal electric gate, where the sheet charge  induces a change in the electric field $\delta E(n_h)$ at the position of the QD. This change $\delta E(n_h)$ depends on the number of stored holes $n_h$. The buildup of the hole gas is a three-step process, starting with the resonant exciton generation (process (1) in Fig.~\ref{f:dev}(c)), followed by hole gas pumping due to field-ionization and tunneling (process (2)) and a depletion by tunneling through the AlAs superlattice (process (3)). 
An increase in the number of holes $n_h$ will blueshift the resonance frequency, and the hole gas population itself is determined by the interplay between processes (2) and (3) in Fig.~\ref{f:dev}(c) with the corresponding pump $\gamma_\textnormal{pump}$ and depletion rate $\gamma_\textnormal{leak}$, respectively. The first step is the resonant generation of electron hole pairs (process (1) in Fig.~\ref{f:dev}(c)) by the laser light, hence, the pump rate $\gamma_\textnormal{pump}$ depends also on the QD resonance $\nu_{QD}$ with respect to the laser frequency $\nu_{L}$, which is the detuning $\Delta \nu=\nu_L - \nu_{QD}$. The resonance position depends again on the electric field across the dot, given by the hole gas that detunes the resonance with respect to the laser frequency. The resulting feedback loop explains the shape and the bistability of the resonance in Fig.~\ref{f:RF}(c) for the situation of "stabilization on" and has been modelled by a rate equation in Merkel et al.~\cite{Merkel.2017}. The width and the asymmetry of the resonance in this voltage range as well as the hysteresis show that the resonance is dragged along as the detuning is shifted. 

The shift of the resonance curve, caused by the quantum-confined Stark effect can be described by $\delta \nu = -p/(h \delta E)$, where $p$ is the excitonic dipole moment and $\delta E$ is the change of electric field. $\delta E$ is given by the hole gas layer, which is approximated as a charged layer inside the heterostructure (Gauss's law):
\begin{equation}\label{e:EFeldShift}
\delta E(n_\T{h}) = - \frac{n_\T{h} e_\T{0}}{A_\T{h} \epsilon_0 \epsilon_\T{r}},
\end{equation}
where $n_h$ is the number of holes, and $A_\T{h}$ is the area covered by the hole gas and is determined by the dimensions of the micropillar (here: $A_\T{h}\approx$ 180\,$\upmu \T{m}^2$). This simple equation already demonstrates that the area of the micropillar structure plays an important role in the hole gas dynamics and can be used to tune and optimize the stabilization feedback loop further in the future. 

The sign on the right-hand side in Eq.~\eqref{e:EFeldShift} is negative, i.~e.~the hole gas layer between the QD and the top gate decreases the electric field. The negative change of electric field translates into a positive change of the resonance frequency: $\nu_\T{QD} \mapsto \nu_\T{QD} + \delta \nu $.  Expanding $\partial n_\T{h}$ around a steady-state hole gas population for small variations of the detuning shows the stabilization behavior, as discussed in Ref.~\cite{Merkel.2017}. For excitation on the low energy side of the resonance, the frequency shift $\delta \nu$ induced by the hole gas  is negatively proportional to a change of the excitonic resonance frequency: $\partial_t(\delta \nu) \propto -\T{d}\nu_\T{QD}$. In other words, when the resonance frequency is shifted by charge fluctuations near the quantum dot, the resulting change of the hole gas population induces an opposite frequency shift of the QD resonance. A direct measurement of this negative feedback is shown in a time-resolved measurement in Fig.~\ref{f:mod} and discussed later. Since the detuning between laser and QD frequency $\Delta \nu$ is directly connected to the number of QD photons, the described negative feedback yields a reduced intensity jitter, i.~e.~a photon noise suppression in the measured resonance fluorescence intensity.

\section{Stabilization of the photon stream}

To switch on the stabilization feedback loop, the detuning is set to a value that is not in the hysteresis region ($\Delta \nu= 3...5 \, \mu$eV in Fig.~\ref{f:RF}(c)), while still achieving a fluorescence count rate as high as possible. 

\begin{figure*}
\centering
\includegraphics[width=1\columnwidth]{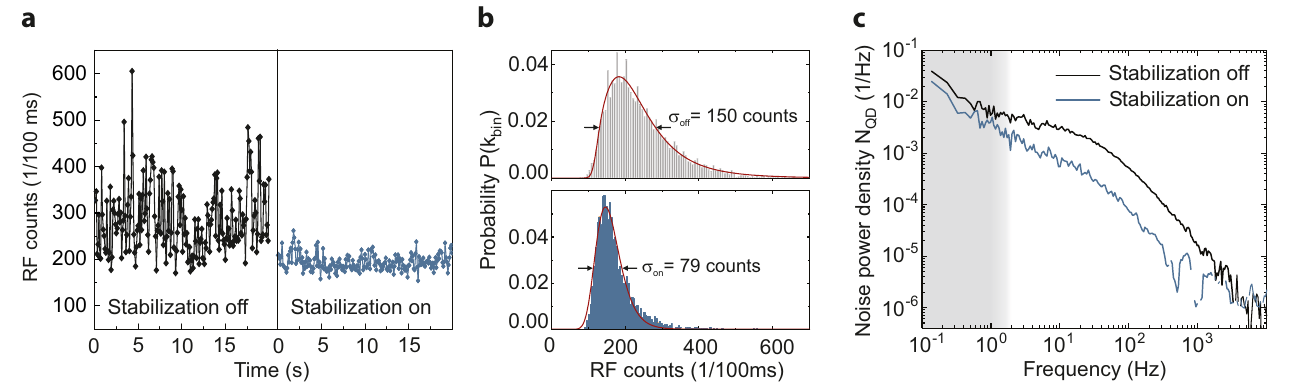}
\caption{Photon noise suppression in the resonance fluorescence (RF) count rate. \textbf{a}, Comparison of two time traces of the RF signal at an integration time of 0.1\,s and a laser power that corresponds to a Rabi frequency of $\Omega_\mathrm{R}=0.4\,\Gamma_0$. The black time trace on the left-hand side has been recorded at a gate voltage outside of the stabilization regime ($V_g=-0.25$\,V), while the blue trace on the right-hand side is measured at $V_g=-0.7$\,V, where the stabilization by the hole gas is "on". \textbf{b}, Histograms of the probability of the RF count rate, where the top and bottom histograms show the count rates without and with stabilization, respectively. The standard derivation $\sigma$ is reduced by almost a factor of two from $\sigma_\textrm{\tiny off}= 150$ counts down to $\sigma_\textrm{\tiny on}= 79$ counts, respectively. \textbf{c}, RF noise spectra of the time traces for hole gas stabilization on (blue line) and off (black line). Plotted is the normalized noise power density of the QD $N_\textrm{\tiny QD}(f)= N_\textrm{\tiny RF}(f)-N_\textrm{\tiny Exp}(f)$, where the Fourier transformed normalized RF $N_\textrm{\tiny RF}(f)$ is corrected by the noise of the experimental setup $N_\textrm{\tiny Exp}(f)$, see Kuhlmann et al.~\cite{Kuhlmann.2013}. The high noise at low frequencies in the gray shaded area is mainly caused by a low RF count rate (shot noise), which cannot be suppressed here by the hole gas stabilization and thus yields only a small reduction of $N_\textrm{\tiny QD}(f)$ after subtraction of $N_\textrm{\tiny Exp}(f)$.}
\label{f:trace}
\end{figure*}

Figure \ref{f:trace}a shows two time-resolved measurements of the resonance fluorescence signal, recorded with a single photon counting setup (see methods below). The black time trace was recorded for a gate voltage of $V_g= -0.35$V outside the hole gas regime ("Stabilization off"), while the blue line at a gate voltage of $V_g= -0.85$V within the hole gas regime ("Stabilization on"). Without the hole gas stabilization, the resonance fluorescence shows large intensity fluctuations in the black time trace. Switching the stabilization on (blue curve), the fluctuations are strongly suppressed, demonstrating the power of this internal feedback loop for photon noise suppression. The average number of RF counts is slightly smaller (by a factor of about 1.3), as the hole gas has to be fed by the exciton generation within the QD, while the the second-order correlation $g^2(t)$ in a Hanbury Brown-Twiss interferometer shows single photon emission with a antibunching dip that falls down to 13 \% (see supplemental information). 

The performance of the stabilization can be evaluated by the histograms in Fig.~\ref{f:trace}(b), where the probability for a certain RF count rate is given for a binning time of 100 ms (time traces of 15\,min). In both recordings, the incident laser power was set to 16\,nW, corresponding to a Rabi frequency of $\Omega_\mathrm{R}=0.4\,\Gamma_0$, where the linewidth (\textit{full width at half maximum}) of this QD is $\Gamma_\mathrm{0} = (4\pm 0.5)\,\mu$eV. The fluctuations in the photon count rate can be roughly quantified by the standard deviation $\sigma_\textrm{\tiny off}= 150$ cts and $\sigma_\textrm{\tiny on}= 79$ cts for stabilization "off" and "on", respectively. This demonstrates a strong suppression in the photon noise by almost a factor of two. 

In order to quantify the stabilization effect on an energy scale, the probability distribution $P(k_\T{bin})$ of $k_\T{bin}$ detections per time bin in Fig.~\ref{f:trace}(b) can be fitted using a photon-counting statistics model as proposed in Ref.~\cite{Matthiesen.2014}:
\begin{align}\label{eq:fullCount}
P(k_\T{bin})	&= \sum_{\Delta \nu}^{} W(\Delta \nu) \: \tilde{P} (k_\T{bin},m(\Delta \nu,\Omega_R)),\\
\T{with } W(\Delta \nu) &= \exp\left[ -\frac{1}{2} \left( \frac{\Delta \nu - \delta_\T{ave}}{\Delta_\T{Diffusion}}\right) 8 
\ln 2 \right].
\end{align}
The model consists of a sum of Poissonian distributions $\tilde{P}$, which depend on the average count rate $m$ per bin $k_\T{bin}$. The average count rate $m$ depends itself on the detuning $\Delta \nu$ and Rabi frequency $\Omega_R$ (given by the laser excitation power). The calculation of $m$ requires the relaxation $T_1=0.8$ ns and dephasing time $T_2=0.33$ ns, which were obtained by linewidth and $g^{(2)}$ autocorrelation measurements (see supplemental information). An undisturbed 2-level system can be described by a single Poissonian distribution. Charge fluctuations influence the detuning between laser and excitonic resonance frequency and, hence, a suitable model consists of a sum of distributions that slightly differ in detuning \cite{Matthiesen.2014}. Every single Poissonian distribution at different detuning is then weighted by a probability factor $W(\Delta \nu)$. The magnitude of the spectral diffusion $\Delta_\T{Diffusion}$ of the underlying Gaussian distribution of $W(\Delta \nu)$ (centered at $\delta_\T{ave}$) is a measure of the strength of the fluctuations. This diffusion constant $\Delta_\T{Diffusion}$ drops to roughly 50\,\% from an initial value of $\Delta_\T{Diffusion}^\T{off}= 220$ MHz down to $\Delta_\T{Diffusion}^\T{on}= 130$ MHz after the stabilization is turned on in Fig.~\ref{f:trace}. 

Figure \ref{f:trace}(c) shows the corresponding noise power spectra. The spectra are calculated from the time traces as described by Kuhlmann et al.\cite{Kuhlmann.2013} (see also supplemental information). Plotted is the normalized noise power density of the QD $N_\textrm{\tiny QD}(f)= N_\textrm{\tiny RF}(f)-N_\textrm{\tiny Exp}(f)$, where the Fourier transformed, normalized RF $N_\textrm{\tiny RF}(f)$ is corrected by the noise of the experimental setup $N_\textrm{\tiny Exp}(f)$. The comparison shows that the hole gas attenuates the fluorescence noise up to a frequency of about 1\,kHz, covering the regime of most of the charge fluctuations \cite{Kuhlmann.2013}. Noise at higher frequencies is predominantly governed by spin noise in the nuclei of the host materials. We presume that the strong noise at low frequencies (grey shaded area in Fig.~\ref{f:trace}) of the spectra is mostly due to our low photon count rates (high shot noise value), where correction by the noise of the experimental setup $N_\textrm{\tiny Exp}(f)$ does not work well. Nevertheless, the overall noise power is reduced by 6 dB by turning on the hole gas stabilization. 

At higher frequencies, the noise spectrum gives an estimate of about how fast the hole gas can react to perturbations. The presented noise comparison suggests time scales of milliseconds. This timescale and the above mentioned noise-suppressing negative feedback can explicitly be probed with a gate modulation experiment, as shown in the next section.

\section{Time-resolved measurements}

\begin{figure}
	\centering
	\includegraphics[width=0.4\columnwidth]{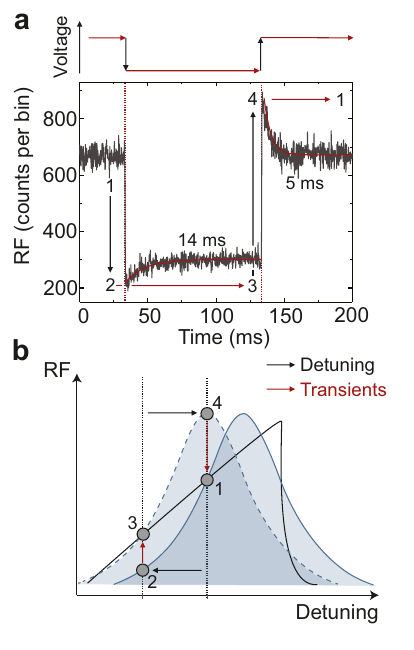}
	\caption{Time-resolved RF measurement of the hole gas stabilization process. \textbf{a}, A gate voltage pulse detunes the excitation laser frequency with respect to the exciton transition (black arrows $\rightarrow$ detuning). The transients (red arrows $\rightarrow$ transients) in the RF counts result from the negative electrical feedback of the hole gas after the abrupt detuning by the gate pulse. It demonstrates the stabilization process in a time-resolved measurement, where a time constant of 5 and 14 ms is observed, respectively. 
	\textbf{b}, Schematic illustration of the stabilization principle, where the blue solid and dashed line show the exciton resonance for two different detunings (given by the applied gate voltage). The solid black line shows the steady-state situation, where the exciton transition is dragged by the hole gas influence, as measured in Fig.~\ref{f:RF}(c). The filled circles at  point 1 and 3 symbolize the measured RF intensity at steady state, while the red arrows symbolize the measured transients from the non-equilibrium situations at point 2 and 4 into equilibrium at point 1 and 3.}
	\label{f:mod}
\end{figure}

A charge fluctuation in the environment of the QD temporarily changes the local electric field and, thus, the detuning between excitonic resonance and laser frequency. A variation of the internal electric field by a gate voltage modulation is equivalent to the effect of a local charge fluctuation, as schematically depicted in Fig.~\ref{f:mod}(a), where black arrows display the detuning by a gate voltage step. As a reaction to a change in detuning, the hole gas settles to a new population. This change in hole gas population can be tracked by measuring the resonance fluorescence, since $n_\T{h}$ is linked to the QD resonance frequency. A time-resolved fluorescence trace is shown in Fig.~\ref{f:mod}(b), obtained by a n-shot measurement \cite{Lu.2010}. A monoexponential fit to the transients is given as red solid line. Two time constants of 14\,ms and 5\,ms are observed for decreasing and increasing voltage, respectively. 

The observed transients and the stabilization behaviour can be easily explained by using a schematic picture of two Lorentzian resonances for an excitonic transition before (solid blue line in Fig.~\ref{f:mod}(b)) and after (dashed blue line) the voltage pulse. The solid black line depicts in Fig.~\ref{f:mod}(b) the steady-state situation with an asymmetric lineshape due to the dragging effect by the hole gas. Therefore, at point (1) and (3) the system "QD--hole gas" is in an equilibrium situation, where the pumping of the hole gas is balanced with its depletion, given by the two involved rates $\gamma_\T{pump}$ and $\gamma_\T{leak}$, respectively. At $t=$ 33\,ms in Fig.~\ref{f:mod}(a), the voltage sets the detuning from a higher to a lower value. The absorption initially decreases at the edge of the resonance (cf.~Fig.~\ref{f:mod}(b) point (2)) and the RF intensity drops sharply (point (1) $\rightarrow$ point (2) in Fig.~\ref{f:mod}(a)) together with the pumping rate $\gamma_\T{pump}$. A decrease in the hole gas pumping yields a decrease of the hole gas population $n_\T{h}$ and a red-shift of the resonance curve towards lower detuning (dashed blue line in Fig.~\ref{f:mod}(b)). While the resonance curve is red-shifted, the RF intensity increases again with the observed time constant of 14\,ms until a new stationary state is reached at point (3). 

At $t=$ 133\,ms, the gate voltage sets the detuning back to a higher value at point (4). The RF intensity increases sharply at this position of the resonance curve and the hole gas pumping rate $\gamma_\T{pump}$ is increased again. With increasing hole gas population, the resonance curve blue-shifts towards higher detuning and the second transient from position (4) to (1) is observed in the RF counts in Fig.~\ref{f:mod}(a). During this blue-shift, the pump rate decreases until the initial stationary state is reached at point (1). 

\section{Discussion}

The experiment in Fig.~\ref{f:mod} simulates an exaggerated frequency fluctuation, by which the negative feedback of the hole gas can be made visible. The time scales of the frequency shift of the resonance curve in reaction to an external perturbation are in the order of milliseconds, in agreement with the attenuation of the photon noise of up to 1\,kHz in Fig.~\ref{f:trace}. The time constant of 5\,ms for the described blue-shift can be controlled by the pumping rate of the hole gas: $\gamma_\T{pump}=\gamma_\T{abs} \, \scL(n_h) \,  P_\T{ionize}$, where the pumping rate itself depends on the Lorentzian lineshape $\scL(n_h)$, the ionization probability of the exciton $P_\T{ionize}$ and the absorption rate $\gamma_\T{abs}$ \cite{Merkel.2017}. As the absorption rate is linked to the laser power, the time constant for stabilization process in the direction of the blue-shift can be decreased by increasing the laser light intensity. A minimal time constant of 2.5\,ms has been measured for a Rabi frequency of $\Omega_\T{R} = 0.4\,\Gamma_0$ for this sample structure with a mesa size of 15\,$\mu$m (see supplemental information). The time constant of 14\,ms for the red-shifted transient in Fig.~\ref{f:mod}(a) is determined by both,  the pumping and depletion rate. The depletion rate $\gamma_\T{leak} = \gamma_\T{esc} n_\T{h}$  can be tuned by the structural properties, where the thickness and height of the tunneling barrier for process (3) in Fig.~\ref{f:dev}(c) are the most obvious.

Besides the electrical detuning and optical incident power control, the stabilization is controlled by the size of the mesa structure. The smaller the micropillar diameter, the larger the change of local electric field per hole, cf.~Eq.~\ref{e:EFeldShift}. Decreasing the lateral size of the micropillar structure decreases the number of holes that are needed to reach an equivalent frequency shift. As a consequence, frequency fluctuations can be compensated by fewer holes and therefore much faster. In the investigated micropillar structure here, the frequency shift per hole is $\delta \nu_{1h}= 0.12$ MHz for a maximum steady state population of about $n_h=4.3\times 10^4$ (see supplemental information). This corresponds to a maximum shift by the hole gas of $\Delta \nu^\T{max}=0.4\,$GHz. Another important parameter for the stabilization performance is the linewidth of the resonance fluorescence curve. Simulations show that smaller linewidths lead to faster stabilization (see supplemental information). This is another advantage of this internal feedback mechanism, since both, small linewidths and high absorption rates, are needed for indistinguishable single photons with high repetition rate. 

\section{Summary}

In summary, we have shown an internal stabilization scheme for the fluorescence of a single self-assembled InAs QD using a two-dimensional hole gas. The hole gas builds up at an interface, located 30\,nm above the QD layer and it is confined by processing the GaAs/AlAs to an electrically contacted mesa structure. The stabilization mechanism is based on a feedback loop between the QD resonance frequency and the hole gas population. A photon-counting statistics model demonstrates a decrease of the distribution width of the fluctuations by 50\,\% (noise power reduction of 6 dB) and the noise spectra show a photon noise reduction for frequencies up to 1\,kHz, directly measured in time-resolved measurements of the resonance florescence signal. This experiment showed the millisecond timescale at which the feedback operates. Higher absorption rates and smaller mesa sizes are expected to yield even higher stabilization speeds (see supplemental information). The most important advantages of the demonstrated internal feedback loop are its easy way of implementation and its scalability to large numbers of photon emitters for optical quantum computation or future quantum networks. 


\section{Methods}
\subsection{Sample and device fabrication}
The sample was fabricated by molecular beam epitaxy on a semi-insulating GaAs(100) substrate, containing a single low density layer of InAs dots (approximately one QD per $\mu$m$^2$) in a GaAs matrix. After a 120\,nm thick GaAs/AlAs superlattice, followed by a 330\,nm thick GaAs layer, the active part of the structure starts with a 30\,nm Si-doped GaAs, which serves as the n-doped back contact and builds up a Schottky diode with a transparent NiCr gate for charge state control. What follows after the Si-doped layer is a tunneling barrier, consisting of 30\,nm undoped GaAs, 10\,nm AlGaAs and 5\,nm GaAs. On top of the GaAs layer, the InAs QDs are grown, followed by 30\,nm GaAs and a 203\,nm thick GaAs/AlAs superlattice as blocking layer. The samples were patterned into micropillar mesa structures by electron beam lithography, were the mesa structures are produced by chemical wet-etching before 6\,nm thick NiCr defines the Schottky top contacts. An Ohmic contact to the Si-doped GaAs was established by deposition of 5\,nm Ni and 230\,nm of AuGe in a thermal evaporator and subsequent annealing at a temperature of 430\,$^\circ$C. Finally, the collection efficiency is enhanced by a half-ball Zirkonia solid immersion lens that was placed on top of the micropillar mesa structure. 

\subsection{Optical measurements}
Resonant optical excitation and collection of the fluorescence light is used to detect the optical response of the single self-assembled QD, where the resonance condition is achieved by applying a specific gate voltage between the gate electrode and the Ohmic back contact. The QD sample is mounted in on a piezo-controlled stage under an objective lens with a numerical aperture of NA$=0.65$, giving a focused spot size with a diameter of about 1\,$\upmu$m. All experiments are carried out in a liquid He confocal dark-field microscope at 4.2\,K with a tunable diode laser for excitation and an avalanche photodiode (APD) for fluorescence detection. The optical path is aligned in photoluminescence measurement (a spectrometer with a 1200-g/mm grating) and the position of the laser spot on the mesa is tracked with a CCD camera. The resonant laser excitation and fluorescence detection is aligned along the same path with a microscope head that contains a 90:10 beam splitter and two polarizers (Thorlabs LPVIS050-MP2). Cross-polarization enables a suppression of the spurious laser scattering into the detection path by a factor of more than $10^7$. For time-resolved measurements, a function generator was used to apply voltage pulses either directly at the gate contact or to a driver of an acoustic-optical modulator (AOM), switching the resonance frequency of the exciton transition or switching the laser beam on/off, respectively. The counts of the APD were binned by a QuTau time-to-digital converter with a temporal resolution of 81\,ps.

\section{Acknowledgments}
This work was supported by the German Research Foundation (DFG) within the Collaborative Research Centre (SFB) 1242 "Non-Equilibrium Dynamics of Condensed Matter in the Time Domain", research project A01 and the individual research grant No.~GE2141/5-1.

\section{Contributions}
A.~A., A.~K.~and M.~G.~planned the experiments. Ar.~L.~
and A.~D.~W.~grew the sample. A. A. and A.~K.~performed the measurements.  A.~A.~and B.~M.~did the simulations. A.~L.~and
M.~G. supervised the work. A.~A.~, A.~L.~and M.~G. wrote the manuscript.

\end{document}